\documentclass[useAMS,usenatbib]{mn2e}

\title[]{Electron-positron bremsstrahlung and pair creation in very high 
magnetic fields}
\author[P. B. Jones]{P. B. Jones\thanks{E-mail:
p.jones1@physics.ox.ac.uk}\\ 
Department of Physics, University of Oxford, Denys Wilkinson Building,\\
Keble Road, Oxford OX1 3RH}

\begin{document}

\date{}

\pagerange{} \pubyear{}

\maketitle

\label{firstpage}

\begin{abstract}
Cross-sections for Rutherford scattering, Coulomb bremsstrahlung and
pair creation, have been calculated at very high magnetic fields in
order to investigate the photo-production of protons at the polar caps
of pulsars whose spin is antiparallel with the polar magnetic flux
density.  The Landau-Pomeranchuk-Migdal effect at very high magnetic
fields is included in a simple electron Green function. 
\end{abstract}


\section{Introduction}
Work on quantum electrodynamic processes at magnetic flux densities of
the order of the critical field $B_{c} = m^{2}c^{3}/e\hbar = 4.41
 \times 10^{13}$ G has been reviewed recently, and very completely, by
Harding \& Lai (2006).  The processes concerned are primarily those
relevant to collisions within plasma at low altitudes above the
surface of a neutron star
and therefore do not include relativistic Coulomb bremsstrahlung
and pair creation
in the interaction of electrons and photons with ions.
There appear to be no published cross-sections for these processes  at
$B \sim B_{c}$ apart from a number of papers on the bremsstrahlung
energy-loss of non-relativistic electrons. But of these, only Lieu (1981)
and Lauer et al (1983) have treated bremsstrahlung as a process of
second order
in the electron-photon coupling and the work of the latter authors is
also
limited to collisions in which both initial and final electrons
are in the lowest Landau state.  Bussard (1980) and Langer (1981) have
considered photon production in non-relativistic electron-ion
plasma collisions but
have treated it as a sequence of first-order processes, that is,
Coulomb
excitation of a higher Landau state followed by cyclotron emission.

However, the
reverse flux of electrons or positrons produced by pair creation in
the open magnetic flux-line region above
neutron-star polar caps must produce electromagnetic showers in the
condensed matter at the surface whose properties are determined
by the relativistic second-order processes. 
In neutron stars with spin
${\bf \Omega}$ such that ${\bf \Omega}\cdot{\bf B} > 0$, the outward
particle flux consists of electrons so that nuclear reactions within
the positron-initiated showers have no obvious observable effect.
But in the
opposite case, ${\bf \Omega}\cdot{\bf B} < 0$, the outward particle
flux is positively charged and, with positrons, also includes protons
that are produced in the shower and diffuse to the surface.  It is
likely that the temporal behaviour of these different components does
lead to instabilities that are the basis of observable phenomena
(Jones 2010). In general, there appears to be no reason a priori why
both ${\bf \Omega}\cdot{\bf B}$ cases should not be present in the
neutron-star population and it is of interest to understand how
each might be observed in the electromagnetic spectrum.
Estimates of shower development in previous work relied
on the correspondence principle, that for large values of the
Landau quantum number the zero-field cross-sections are valid, but
this is unlikely to 
be satisfactory at $B \sim B_{c}$ where the number of contributing
Landau states can be small.
The present paper attempts to give cross-sections
that are rather better than order-of-magnitude estimates but are
not of high accuracy owing to the discrete nature of the Landau
spectrum and to truncations that are made in performing
the calculations.
Its purpose is to see whether and in what way the cross-sections differ 
qualitatively from those at zero field.
Nucleon production in electromagnetic showers occurs almost entirely
through formation and decay of the giant dipole resonance.  Thus the
paper is concerned, particularly, with factors that influence
the low-energy photon
track length and its depth distribution.

With neglect of radiative corrections and of the anomalous magnetic
moment, the energy states of an electron in a uniform magnetic field are
\begin{eqnarray}
E = \pm \sqrt{p^{2} + 1 + 2nB}
\end{eqnarray}
in which $p$ is the longitudinal momentum component,
parallel with ${\bf B}$, and
$n = 0, 1, 2,...$ is the Landau quantum number.  Apart from the
Appendix and where explicitly stated otherwise,
momentum and energy are here expressed in units of $mc$ and
$mc^{2}$; the magnetic flux density $B$ is in units of $B_{c}$.
In electron-photon interactions, the conserved quantities are energy
and the longitudinal momentum component.  Thus the kinematic
behaviour of an electron is that of a particle confined
to one dimension and of mass $m\sqrt{1 + p_{\perp}^{2}}$ in which
we have defined, purely for ease of notation, a notional transverse
momentum component $p_{\perp} = \sqrt{2nB}$.

We have adopted the solutions of the Dirac equation found by
Johnson \& Lippmann (1949) and give a summary of calculational details
in the Appendix.  These are unremarkable except that owing to the
one-dimensional kinematics of electrons and the small number of Landau
states that contribute significantly at $B\sim B_{c}$, we have found
it convenient, given that numerical computation must be involved,
not to proceed with the usual approach of
covariant formalism and Green functions but instead to write down
the transition matrix to second order directly in terms of
time-ordered matrix elements
calculated using an explicit representation of the Dirac matrices.
It is also appropriate to mention here that the states given by
equation (1) are two-fold degenerate for $n > 0$ only if
radiative level shifts, natural widths, and the anomalous magnetic
moment of the electron are neglected.  But for the high-energy
spin-averaged or summed processes considered here, the fine
structure is of no consequence and we use only the completeness
of the Johnson-Lippmann eigenfunctions.  References and
some further details are given in the Appendix.

The most simple problem here is that of finding the
replacement for
Rutherford's scattering formula. Small-angle Coulomb scattering
in the zero-field case is replaced at $B \sim B_{c}$ by the Coulomb
excitation of Landau states.  This has been treated by Bussard (1980)
and Langer (1981) but only in the non-relativistic limit.  The
relativistic cross-sections that we require
appear not to have been
published previously and are given in Section 2.  There have been many
calculations of transition rates for cyclotron emission, and a summary
is contained in the review of Harding \& Lai (2006).  New transition
rates are given in Section 3 to allow comparison with previous work
and, more specifically, to obtain the distribution of photon transverse
momentum $k_{\perp}$ from the decay of $n > 1$ Landau states which
is of crucial importance in the development of
electromagnetic showers at $B \sim B_{c}$.
Cross-sections for Coulomb pair creation and bremsstrahlung are
obtained in
Sections 4 and 5. In the case of pair creation, there appears to be
no previously published work.  For bremsstrahlung, the work of
Lieu (1981) and Lauer et al (1983) is in the non-relativistic limit
so that comparison is not possible.
There is also the further problem
that calculations relying on truncation of the number of Landau states
lack external tests of their correctness because the zero-field
limit is not available.  But where minor tests of correctness exist, we
find that they are satisfied.  Finally, with primary electron energies
of the order of $10^{3}$ GeV and the high condensed matter densities
predicted at the neutron star surface, it is essential to consider the
Landau-Pomeranchuk-Migdal effect and its influence on Coulomb pair
creation and bremsstrahlung cross-sections.  Its existence and broad 
properties at zero field have been verified through experiments at
particle accelerators.  We refer, in particular, to the recent work
of Hansen et al (2004) and to the review of Klein (1999).  We find
that the effect exists at $B \sim B_{c}$ and that the one-dimensional
electron kinematics makes possible a particularly simple treatment
which is given in Section 6. The cross-section calculations of
Sections 4 \& 5 have been repeated with allowance for the LPM effect
which is shown to significantly change early shower development.
The final section gives a qualitative summary of the cross-sections
obtained and of
the effect of magnetic fields of the order of the critical field on
shower development.

\section[]{Rutherford scattering}

Small-angle Coulomb scattering at zero field, in which the
transverse momentum component is a
continuous variable, is replaced by a Landau transition
$n \rightarrow n^{\prime}$ with longitudinal moment transfer $q$.
Free atoms in high magnetic fields have an axially symmetric but
complex
electron density distribution.  But the departures from uniform
electron density in condensed matter are not large except for a
small number of inner orbitals
and we therefore adopt the potential $Ze\tilde{V}$ of a point
charge within a spherical
Wigner-Seitz unit cell of uniform electron density.
Its radius is
\begin{eqnarray}
r_{WS} = 2.15\times 10^{-10}Z^{0.23}B^{-0.40} \hspace{1cm} {\rm cm},
\end{eqnarray}
obtained from the ion number density calculated by Medin \& Lai (2006)
for atomic number $Z$.  Calculational details are summarized in
the Appendix.  The symmetry of $\tilde{V}$ indicates that Landau
functions in cylindrical polar coordinates are the appropriate choice.
The quantum number $l$ denotes the spatial degeneracy of these
states and the procedure is to average the transition
rate over all $l$ in the range $0 \le l \le l_{m}$.
Each state has a guiding centre radius $\sqrt{2l + 1}r_{B}$,
where $r_{B}$ is the cyclotron radius.
The incident flux is then $c$ divided by $(2l_{m} + 1)\pi r_{B}^{2}$,
and the cross-section at incident momentum $p \gg 1$,
with no spin-flip, is the limit $l_{m}\rightarrow \infty$ of,
\begin{eqnarray}
\lefteqn{\sigma _{nn^{\prime}} = \pi r_{B}^{2}\left(\frac{Ze^{2}}{\hbar 
c}\right)^{2}
\left(\frac{2l_{m} + 1}{l_{m} + 1}\right)}   \hspace{3cm}  \nonumber \\
\sum^{l_{m}}_{l=0}
\left|\langle p+q, n^{\prime}, l^{\prime}|\tilde{V}
|p, n, l\rangle\right|^{2}.
\end{eqnarray}
This expression is greatly simplified by the selection rule
$\delta(n - l) = 0$ arising from equation (A3) and from the axial
symmetry of $\tilde{V}$.
Spin-flip cross-sections are smaller by a factor of the order of
$(p_{\perp}^{\prime} - p_{\perp})^{2}/4E^{2}$ relative to unity and
as our interest is primarily in electrons with $p \gg mc $ they
have not been calculated.  The range of initial-state quantum numbers
$l$ should be so large that the cross-section is independent
of $l_{m}$ in the case of an isolated atom.  We have confirmed 
that, for $B = B_{c}$, this condition is satisfied adequately for
$l_{m} \approx 30$ at which value guiding centres are at radii
approaching the Wigner-Seitz cell radius.
Numerical cross-sections $\sigma _{nn^{\prime}}$ are given in
Table 1, for $n, n^{\prime} \leq 4$.  They are in units of
$1 {\rm bn} = 10^{-24}$ cm$^{2}$
and are for $Z = 26$  and longitudinal momentum $p = 20$,
but are almost exactly
independent of $p$.  Calculation to the lowest order in $\tilde{V}$
is satisfactory here, as for the zero-field small-angle 
Rutherford formula,
owing to the pole in the amplitude at zero momentum transfer. Thus
the cross-sections are linear in $Z^{2}$.

The cross-sections of Table 1 are valid in the relativistic limit
of $p\gg 1$ so that comparison with the near-threshold
cross-sections found by Bussard (1980) and Langer (1981) is not
possible. But the standard Rutherford cross-section at
small angles can be expressed in terms of a continuous variable
$n$ through the
correspondence principle relation $p_{\perp} =\sqrt{2nB}$.  It is
$d\sigma _{R}/dn = 0.50Z^{2}/n^{2}B$ bn.  The numerical values and
dependences on $n$ and $B$ are quite close to those given in the
Table for an initial state with $n = 0$.  The $n \rightarrow n$
cross-sections are well-defined by equation (3) even
though they are to an evanescent state whose asymptotic final
form with $q = 0$ is indistinguishable from the
unscattered.  These $\delta n = 0$ transitions to virtual
states of finite $q$ appear
in the calculation of second-order processes, in
particular bremsstrahlung and pair creation
cross-sections, and in condensed matter are relevant to the
Landau-Pomeranchuk-Migdal effect considered in Section 6.

The cross-sections in Table 1 are typically two orders of
magnitude larger than those we obtain for Coulomb bremsstrahlung
in Section 5.  Thus Rutherford 
scattering excites higher Landau states whose decay is
an important source of photons, particularly in the later stages
of shower development.  We shall refer to this in greater
detail in Section 7.

\begin{table}
\caption{Cross-sections in units of $10^{-24}$ cm$^{2}$
for Coulomb-induced Landau
transitions $n \rightarrow n^{\prime}$, with $n, n^{\prime} \leq 4$,
are almost exactly independent of longitudinal momentum
at $p\gg mc$ but have been calculated
for electrons of $p =20$ mc incident on
an isolated neutral atom of atomic number $Z = 26$.
The magnetic flux
density in the first column is in units of
$B_{c} = 4.41\times 10^{13}$ G.  The significance of the
$n \rightarrow n$ transitions is considered in the text and later
in Section 6.}

\begin{tabular}{@{}llrrrrr@{}}
\hline
$B$ &  $n$  & $n^{\prime}=0$ &  1  &  2  &  3  &  4    \\
\hline
0.3  &  4  &  91  &  271  &  1132  &  6128  &  20886  \\
     &  3  &  182  &  841  &  5454  &  23132  &  6146  \\
     &  2  &  525  &  4483  &  25901  &  5467  & 1138  \\
	 &  1  &  3032  &  29470  &  4491  & 843  &  272  \\
	 &  0  &  34574  &  3036  &  526  &  182  &  91  \\
\hline
1.0  &  4  &  27  &  80  &  334  &  2056  &  8729   \\
     &  3  &  54  &  248  &  1804  &  9543  &  2063  \\
	 &  2  &  156  &  1460  &  10521  &  1810  &  336   \\
	 &  1  &  968  &  11746  &  1463  &  249  &  80   \\
	 &  0  &  13439  &  970  &  156  &  54  &  27   \\
\hline
3.0  &  4  &  9  &  25  &  102  &  713  &  3862   \\
     &  3  &  17  &  77  &  621  &  4179  &  716  \\
	 &  2  &  49  &  498  &  4554  &  623  &  103  \\
	 &  1  &  328  &  5012  &  500  &  78  &  25   \\
	 &  0  &  5628 &  329  &  49  &  17  &  9   \\
\hline

\end{tabular}
\end{table}

\section[]{Cyclotron emission}

Transition rates for relativistic cyclotron emission are obviously
relevant to the low-altitude plasma above polar caps at $B \sim B_{c}$
and therefore have been
calculated by a number of authors
(Herold, Ruder \& Wunner 1982,
Latal 1986, Baring, Gonthier \& Harding 2005; see also the review of
Harding \& Lai 2006).  The principal reason for our calculation here,
which is limited to unpolarized electrons, is to investigate
the sequence of partial transition rates by which an electron in
Landau state $n$ decays to $n = 0$.  These determine the distribution
of the transverse photon momentum $k_{\perp} = k\sin\theta$, where
$\theta$ is the photon angle with respect to ${\bf B}$, and hence
electromagnetic shower development at $B \sim B_{c}$. 

In the rest frame of the initial electron, the relation between  
momentum $k$ and $\theta$ is,
\begin{eqnarray}
k\sin^{2}\theta = E_{0} - \left(E_{0}^{2} -\sin^{2}\theta
(p_{\perp}^{2} - p_{\perp}^{\prime 2})\right)^{1/2},
\end{eqnarray}
in which $E_{0} = \sqrt{1 + p_{\perp}^{2}}$.  The partial transition
rate for unpolarized electrons is
\begin{eqnarray}
\Gamma _{nn^{\prime}} = \frac{1}{\hbar}\int^{1}_{-1}d(\cos\theta)
\frac{k^{2}}{2\pi\hbar^{3}c}
\frac{E^{\prime}}{E^{\prime} + ck\cos^{2}\theta}   \nonumber  \\
\frac{1}{2(l_{m}+1)}\sum _{\epsilon}\sum _{s,s^{\prime}}
\sum _{l,l^{\prime}=0}^{l_{m},l^{\prime}_{m}}
\left|\langle l^{\prime},n^{\prime},s^{\prime},{\bf 
k},\epsilon|e\mbox{\boldmath$\alpha$}
\cdot\tilde{\bf A}|n,l,s\rangle \right|^{2},
\end{eqnarray}
summed over photon polarization states $\epsilon _{\parallel}$ and
$\epsilon _{\perp}$, respectively parallel with and perpendicular
to ${\bf k}\times{\bf B}$, and over final electron spins.
The radiation field vector potential is $\tilde{\bf A}$ given by
equation (A6) and the final-state electron energy is $E^{\prime}$.
The transition rate is expressed here as an average over
$l_{m} + 1$ spatially degenerate initial states and we have verified
that the computed rates are $l_{m}$-independent, as
they should be, and have used the values $l_{m} = 0$ and
$l_{m}^{\prime} = 12$.
The computed values, in units of
$10^{-3}\omega _{B}$, where $\omega _{B}$ is the cyclotron angular
frequency, are given in Table 2 for $n\leq 4$. The summed
transition rates given in the final row of each section are the sole
results of this paper that can be compared directly with previously
published work.  Agreement with the rates shown
in Fig. 3 of Harding \& Lai (2006) is satisfactory insofar as
accurate comparison is possible.
It is interesting to see that there is effectively no selection
rule on $\delta n$ and that the partial rates display the
$\delta n$-dependence noted by these authors in that, with
increasing magnetic flux density, there appears an enhanced
rate for direct transitions  to the $n = 0$ ground state.  The
dependence on photon polarization is not given here.  It is
unremarkable: transitions to $\epsilon _{\parallel}$
are approximately twice as strong as those to the $\epsilon _{\perp}$.
Photon angular distributions vary with $\delta n$ but are
also unremarkable and are not given here.
For $\delta n = -4$ they have a maximum at $\theta = \pi/2$; for
$\delta n = -1$ the maxima are near $\theta = 0, \pi$.

The maximum photon transverse momentum occurs at $\theta = \pi/2$
and is,
\begin{eqnarray}
k_{\perp,max} = \sqrt{1 + p_{\perp}^{2}} - \sqrt{1 + p_{\perp}^{\prime 2}}.
\end{eqnarray}
It is not possible to make a compact statement about the
distributions of $k_{\perp}$ because the values of $k_{\perp,max}$
are functions of both $n$ and $B$.  The critical thresholds
for single-photon magnetic pair creation are those at
$k_{\perp} = 2$ for electron-positron Landau numbers
$n_{\pm} = n_{\mp} = 0$ and
$k_{\perp} = 1 + \sqrt{1 + 2B}$ for $n_{\pm} = 0$ with $n_{\mp} = 1$.
Rates immediately above the lower threshold are partially
inhibited by a selection rule but above the higher threshold,
transition rates for magnetic pair creation are so high that
photon mean free paths are small compared with the radiation length,
which is the basic unit of length for shower development.  But
Table 2 shows that, except for the highest magnetic field, the
majority of transitions from $n \leq 4$ have $k_{\perp} < 2$ and so
are not followed by magnetic pair creation.  Thus cyclotron emission,
following excitation by Rutherford scattering, is an important
source of photon track length in a shower.

\begin{table}
\caption{Rates for cyclotron emission with Landau transitions
$n \rightarrow n^{\prime}$ are given in units of
$1.0\times 10^{-3}\omega _{B}$ for unpolarized electrons.
The sums of the partial transition rates for
$n \leq 4$ are in the final row of each section.}

\begin{tabular}{@{}rcrrrr@{}}
\hline
$B$  &  $n^{\prime}$ &  $n = 1$  &  2  &  3  &  4   \\
\hline
0.3  &  3  &        &        &        &  1.33    \\
	 &  2  &        &        &  1.33  &  0.68   \\
	 &  1  &        &  1.24  &  0.59  &  0.38   \\
	 &  0  &  0.76  &  0.39  &  0.26  &  0.20    \\
	 & sum &  0.76  &  1.63  &  2.18  &  2.59    \\
\hline
1.0  &  3  &        &        &        &  0.97    \\
     &  2  &        &        &  1.05  &  0.59    \\
	 &  1  &        &  1.14  &  0.61  &  0.42    \\
	 &  0  &  1.06  &  0.67  &  0.52  &  0.45    \\
	 & sum &  1.06  &  1.81  &  2.18  &  2.43    \\
\hline
3.0  &  3  &        &        &        &  0.61      \\
     &  2  &        &        &  0.68  &  0.39    \\
	 &  1  &        &  0.79  &  0.43  &  0.30    \\
	 &  0  &  0.94  &  0.62  &  0.50  &  0.44    \\
	 & sum &  0.94  &  1.41  &  1.61  &  1.74    \\
\hline
10.0 &  3  &        &        &        &  0.35    \\
     &  2  &        &        &  0.39  &  0.22    \\
	 &  1  &        &  0.46  &  0.25  &  0.17    \\
	 &  0  &  0.62  &  0.41  &  0.34  &  0.29   \\
	 & sum &  0.62  &  0.87  &  0.98  &  1.03   \\
\hline

\end{tabular}
\end{table}

\section[]{Pair creation}

Coulomb pair creation in a uniform magnetic field requires
a longitudinal momentum transfer $q$ from the nucleus,
\begin{eqnarray}
q = k(1 - \cos\theta) - \frac{1}{2p_{+}}
\left(1 + p_{\perp+}^{2}\right)
 - \frac{1}{2p_{-}}\left(1 + p_{\perp-}^{2}\right),
\end{eqnarray}
an expression valid for final-state electron and positron momenta
$p_{\pm} \gg 1$, and therefore differs from the zero-field
case in being a function
of both $k$ and $k_{\perp}$.  The momentum transfer is a minimum
for $p_{+} = p_{-}$ and $p_{\perp+} = p_{\perp-} =0$ for which
conditions it falls to zero at $k_{\perp} = 2$, the kinematic
threshold for magnetic pair production.
The transition matrix element to second order for an isolated atom is,
\begin{eqnarray}
M_{\parallel,\perp}^{P} =
 \langle p_{-},n_{-},l_{-},s_{-}|-Ze^{2}\tilde{V}|
p_{-}-q,n_{-},l_{-},s_{-}\rangle    \nonumber   \\ 
\left(E_{-}(p_{-}) - E_{-}(p_{-} - q) +i\eta\right)^{-1}
  \nonumber \\
\langle p_{-}-q,n_{-},l_{-},s_{-}|e\mbox{\boldmath$\alpha$}\cdot
\tilde{\bf A}|-p_{+},n_{+},l_{+},-s_{+}\rangle  \nonumber  \\
+ \langle p_{+},n_{+},l_{+},s_{+}|Ze^{2}\tilde{V}|
p_{+}-q,n_{+},l_{+},s_{+}\rangle    \nonumber \\
\left(E_{+}(p_{+}) - E_{+}(p_{+}-q) + i\eta\right)^{-1} \nonumber \\
\langle p_{-},n_{-},l_{-},s_{-}|e\mbox{\boldmath$\alpha$}\cdot
\tilde{\bf A}|-p_{+}+q,n_{+},l_{+},-s_{+}\rangle
\end{eqnarray}
in which $\eta > 0$ is infinitesimal. The subscripts $\parallel$
and $\perp$ refer to the photon polarization with respect to
${\bf k}\times{\bf B}$.
The spin and spatial degeneracy states are
labelled by $s_{\pm}$ and $l_{\pm}$ (see equations A7 and A8).
There are two further
time-ordered terms in the amplitude but they involve virtual
pair creation in the Coulomb field.  They have energy denominators
that are large, of the order of $k$ rather than $k^{-1}$ as in the
above, and for $k\gg 1$ they can be neglected.  The total
cross-section is then,
\begin{eqnarray}
\sigma^{P} _{\parallel,\perp} = Z^{2}\left
(\frac{e^{2}}{\hbar c}\right)^{3}\left(\frac{\hbar}{mc}\right)^{2}
\int^{k-\sqrt{1+p_{\perp+}^{2}}}_{\sqrt
{1+p_{\perp-}^{2}}}\frac{dE_{-}}{2\pi}
\frac{E_{-}E_{+}}{p_{-}p_{+}} \nonumber  \\
\sum _{s_{\pm}}\sum _{n_{\pm}=0}^{n_{m}}\sum _{l_{\pm}=0}^{l_{m}}
\left|M^{P}_{\parallel,\perp}\right|^{2}
\end{eqnarray}
for pair creation by a photon of momentum $k$, transverse
momentum $k_{\perp}$, and of polarization
 $\mbox{\boldmath$\epsilon _{\parallel}$}$ or
 $\mbox{\boldmath$\epsilon _{\perp}$}$. 
In this expression,
the charges have been factored out from the matrix element (8).
There are a number of approximations here that merit some
consideration. Firstly, truncation of the
Landau quantum numbers to $n_{\pm} \leq 4$ has been tested
by examination of the computed cross-section as a function of
$n_{m}$ for $0 \leq n_{m} \leq 4$.  The basis for this
truncation is
that, for fixed $k$, the amplitude given by equation (8) is
approximately
$\propto p^{-1}_{\perp\pm}$, also that matrix elements tend
rapidly to
zero as they are constructed between states with increasingly
different quantum numbers and hence nodal surfaces. 
There is evidence of reasonably satisfactory convergence, but
cross-sections obtained here should be seen as lower limits.
Truncation of the sum over the spatial degeneracy quantum numbers
to $l_{\pm} \leq 12$ again produces fair convergence.  But in
this case, there is a natural upper limit determined by the
radius of the Wigner-Seitz cell.  Finally, the matrix elements
of the Coulomb potential $Ze\tilde{V}$ have been limited to
those with $\delta n = 0$.  This is satisfactory, for the
purposes of the present work, because these are typically
much larger than off-diagonal matrix elements, as shown by the
cross-sections of Table 1.  But errors arising from this latter
truncation are not positive-definite.

Numerical values of the isolated-atom total cross-sections are
given in Table 3
for ranges of values of $B$, $k$ and $k_{\perp}$ and for both
photon polarizations.  They are for $Z = 26$ but being of
second order in electron-photon coupling, are linear in
$Z^{2}$.
The differential cross-sections are unremarkable
symmetric functions of the final-state energies $E_{\pm}$,
not strongly peaked within the kinematic
limits present in the integral of equation (9).
They are not given here primarily because in the relevant case of
condensed matter they may be unpredictably distorted.
The explanation for this is a consequence of the partial order
which may be present in neutron-star surface
condensed matter, and is deferred until Section 7.

\begin{table}
\caption{Numerical values of the total cross-section
(in units of $1{\rm bn} = 10^{-24}$ cm$^{2}$) for
pair creation in the Coulomb field of an isolated 
atom with nuclear charge $Z = 26$ are given for both
photon polarizations in columns 3-5 of the upper sector of
the Table.  The
photon momentum $k$ and transverse momentum $k_{\perp}$
are in units of $mc$.  The magnetic flux density is in units
of the critical field $B_{c} = 4.41\times 10^{13}$ G. The
three-row lower sector gives cross-sections re-calculated
in condensed
matter with inclusion of the Landau-Pomeranchuk-Migdal effect.
We refer to Section 6 for the details of this latter calculation,
which assumes a static structure function $S(q) = 1$.}

\begin{tabular}{@{}rrrrr@{}}
\hline
  $B$ & $k$ &  & $\sigma^{P} _{\parallel},\sigma^{P} _{\perp}$ & \\
\hline
    &    &  $k_{\perp} = 0.5$  &  1.0  &  1.5     \\
\hline
  0.3  &    40  &  2.09, 2.03  &  4.21, 4.20  &  4.39, 4.60  \\
       &   400  &  2.28, 2.21  &  4.55, 4.53  &  4.70, 4.91 \\
	   &  4000  &  2.30, 2.23  &  4.59, 4.57  &  4.73, 4.95  \\
  1.0  &    40  &  0.61, 0.59  &  1.22, 1.23  &  2.00, 2.52  \\
       &   400  &  0.71, 0.69  &  1.41, 1.43  &  2.30, 2.87   \\
	   &  4000  &  0.72, 0.70  &  1.44, 1.49  &  2.33, 2.90   \\
	   & 40000  &  0.73, 0.70  &  1.44, 1.45  &  2.34, 2.91   \\
  3.0  &    40  &  0.19, 0.19  &  0.29, 0.31  &  0.44, 0.87    \\
       &   400  &  0.24, 0.24  &  0.37, 0.40  &  0.57, 1.11   \\
	   &  4000  &  0.25, 0.25  &  0.38, 0.41  &  0.59, 1.13    \\
 10.0  &    40  &  0.06, 0.06  &  0.07, 0.08  &  0.08, 0.23    \\
       &   400  &  0.07, 0.07  &  0.08, 0.10  &  0.11, 0.35   \\
	   &  4000  &  0.07, 0.07  &  0.09, 0.11  &  0.11, 0.36  \\
\hline
  1.0  &    40  &  0.61, 0.59  &  1.22, 1.23  &  1.99, 2.47  \\
       &   4000 &  0.21, 0.21  &  0.51, 0.50  &  0.78, 0.79  \\
	   &  40000 &  0.02, 0.02  &  0.05, 0.05  &  0.08, 0.08  \\
\hline
  
\end{tabular}
\end{table}

A problem in this Section is that there are no external tests
of the validity of expressions
derived from equations (8) and (9) or of their numerical
evaluation.  But we note that the differential cross-sections
are symmetric in $E_{\pm}$ and that under the restriction
$n_{m} = 0$, the cross-section is $\sigma^{P} _{\parallel} = 0$
for photons polarized parallel with ${\bf k}\times{\bf B}$.
This is consistent with the fact that for this polarization,
the matrix element (A7) should be exactly zero for
$n_{+} = n_{-} = 0$ (see Semionova \& Leahy 2001).

The cross-sections in Table 3 are quite slowly varying functions
of $k$  and can be compared with the
asymptotic $k \gg 1$ Bethe-Heitler value of
$\sigma ^{P} = 5.0$ bn for unpolarized photons
incident on a screened $Z = 26$ nucleus in zero magnetic field.  
The correspondence principle indicates that the Bethe-Heitler
expression with modified screening should be a fair approximation
at moderate fields,
of the order of $10^{12}$ G, for which the Landau quantum number
associated with $p_{\perp}\sim 1$ is large.  It is difficult
to obtain cross-sections for fields between $10^{12}$ G and
those of the Table because the truncations we have used are
unlikely to be satisfactory approximations.  But as a function
of decreasing $B$, the cross-sections in the Table are not
inconsistent with smooth convergence at intermediate fields
toward the Bethe-Heitler value. The most noticeable features of
the Table are: (i) the rapid increase of cross-section as $k_{\perp}$
approaches the $n_{\pm}$-dependent thresholds for magnetic
pair production at which $q$ and hence the energy denominators
in equation (8) become zero; and (ii), the rapid decrease of
cross-sections at $B > B_{c}$.

\section[]{Bremsstrahlung}

The emission of a photon of momentum ${\bf k}$ at an angle $\theta$
with ${\bf B}$, and a change of Landau quantum number
$n \rightarrow n^{\prime}$, requires a longitudinal momentum
transfer from the atom of
\begin{eqnarray}
q=-k(1-\cos\theta) +\frac{1+p_{\perp}^{2}}{2p} - 
\frac{1+p^{\prime 2}_{\perp}}{2p^{\prime}},
\end{eqnarray}
where the initial electron longitudinal momentum is $p \gg mc$
and the final momentum is $p^{\prime} = p + q - k\cos\theta$.

The transition matrix element, to second order for an isolated
atom, is
\begin{eqnarray}
M^{B}_{\parallel,\perp} = \langle p^{\prime},
{\bf k}, n^{\prime}, l^{\prime},s^{\prime}|
e\mbox{\boldmath$\alpha$}\cdot
\tilde{\bf A}|p+q,n,l,s\rangle  \nonumber  \\
(E(p) - E(p+q) + i\eta)^{-1}    \nonumber   \\
\langle p+q,n,l,s|-Ze^{2}\tilde{V}|p,n,l,s\rangle + \nonumber  \\
\langle p^{\prime},n^{\prime},l^{\prime},s^{\prime}|
-Ze^{2}\tilde{V}|p^{\prime}-q,n^{\prime},l^{\prime},
s^{\prime}\rangle    \nonumber   \\
(E(p^{\prime}) - E(p^{\prime} - q) + i\eta)^{-1} \nonumber  \\
\langle p^{\prime} - q,{\bf k}, n^{\prime},
 l^{\prime},s^{\prime}|e\mbox{\boldmath$\alpha$}\cdot
 \tilde{\bf A}|p, n, l,s\rangle,
\end{eqnarray}
in which the subscripts $\parallel$ and $\perp$ again refer to the
photon polarization with respect to ${\bf k}\times{\bf B}$.  As
in the previous Section, and for the same reason,
we have neglected two time-ordered terms in the amplitude
that involve virtual pair creation in the Coulomb
field. The
existence of single-photon magnetic pair creation thresholds
implies that
the bremsstrahlung cross-section in a high magnetic field is
best expressed in terms of the independent variables $k$ and
$k_{\perp}$.  The cross-section defined by the energy-loss rate
(equivalent to the radiation
length) is then given by an integration over photon energy,
\begin{eqnarray}
\sigma^{Rad} = \int^{k_{m}}_{0}dk\frac{k}{p}\frac{d\sigma^{B}}{dk},
\end{eqnarray}
in which the differential bremsstrahlung cross section for
an electron in Landau state n,
summed over both photon polarization states, is,
\begin{eqnarray}
\frac{d\sigma^{B}}{dk} = (2l_{m} + 1)r^{2}_{B}
Z^{2}\left(\frac{e^{2}}{\hbar c}\right)^{3}
\left(\frac{1 + \delta _{n0}}{4}\right)\int\frac{k_{\perp}}
{k}dk_{\perp}     \nonumber    \\
\sum _{l,l^{\prime}=0}^{l_{m}}
\sum _{\epsilon}
\sum _{s,s^{\prime}}\sum _{n^{\prime}=0}^{n_{m}}
\left|M^{B}_{\epsilon}\right|^{2},
\end{eqnarray}
The summations have been truncated to $n_{m} = 4$ and
$l_{m} = 12$, as in the case of pair creation, but this
approximation is much less satisfactory for bremsstrahlung and
has lead us to omit values for $B < B_{c}$.  Bearing in mind
the importance of the variable $k_{\perp}$, the 
cross-section $\sigma^{Rad}$ given by equations (12) and (13) has
been divided into two components: $\sigma^{Rad}_{1}$ is for the
interval $0 < k_{\perp} < 2.5$ and $\sigma^{Rad}_{2}$ for
$2.5 < k_{\perp} < 5.0$ mc.
The computed values in units of $10^{-24}$ cm$^{2}$, for
electrons in an initial $n = 0$ Landau state incident on isolated
atoms are given in the upper sector of Table 4.  They are summed
over both photon polarizations, but we note that the
$\epsilon _{\perp}$ cross-section is typically an order of
magnitude larger than that for $\epsilon _{\parallel}$. 
The computed values are almost completely independent of
$p$ and there is little cross-section outside the intervals of
$k_{\perp}$ considered.  The total cross-section is simply
defined here as the sum of $\sigma^{Rad}_{1,2}$.  Cross-sections
for initial Landau states $n > 0$ are of similar magnitude
but are not given here.  These initial states are, of course,
naturally unstable against photon decay, which appears as a
zero in the denominators of equation (11) and is discussed
further in Section 7.

\begin{table}
\caption{Numerical values of the cross-section equivalent to the
radiation length, defined by equation (12), are given in the upper
sector of the Table for electrons in the $n = 0$ Landau state
incident on
isolated atoms of $Z = 26$. The values of electron momentum $p$
are in units of mc, and $B$ is in units of the critical field
$B_{c} = 4.41 \times 10^{13}$ G.  Columns 3 and 4 are partial
cross-sections defined for the photon transverse momentum
intervals $0 < k_{\perp} < 2.5$ and $2.5 < k_{\perp} < 5.0$ mc,
respectively. The total cross-section in column 5 is defined here as
the sum of these partial cross-sections.  The 15 rows in the lower
sector of the table give the above cross-sections re-calculated
in condensed matter assuming a static structure function $S(q) = 1$
but with inclusion of the Landau-Pomeranchuk-Migdal
effect.  We refer to Section 6 for further details.}

\begin{tabular}{@{}rrrrrr@{}}
\hline
  p  &  B  &  $\sigma^{Rad}_{1}$ & 
     $\sigma^{Rad}_{2}$ &   $\sigma^{Rad}$    \\
\hline
 200  &  1.0  &  4.83  &  1.85  &  6.68   \\
      &  3.0  &  0.77  &  1.44  &  2.21   \\
	  & 10.0  &  0.08  &  0.25  &  0.33      \\
2000  &  1.0  &  4.82  &  1.85  &  6.67     \\
      &  3.0  &  0.77  &  1.44  &  2.21      \\
	  & 10.0  &  0.08  &  0.25  &  0.33   \\
20000 &  1.0  &  4.82  &  1.85  &  6.67    \\
	  &  3.0  &  0.77  &  1.44  &  2.21    \\
	  & 10.0  &  0.08  &  0.25  &  0.33    \\
\hline
    200  &  1.0  &  22.83  &  1.87  &  24.70  \\
	     &  3.0  &   9.94  &  1.63  &  11.54   \\
		 & 10.0  &   3.92  &  0.76  &  4.68   \\
   2000  &  1.0  &  20.04  &  2.41  &  22.45  \\
         &  3.0  &   6.95  &  3.59  &  10.54   \\
		 & 10.0  &   1.75  &  1.88  &  3.63   \\
  20000  &  1.0  &   5.35  &  2.15  &  7.50   \\
         &  3.0  &   1.32  &  2.66  &  3.98   \\
		 & 10.0  &  0.22   &  0.78  &  1.00   \\
  60000  &  1.0  & 1.52  &  0.95   &  2.47    \\
         &  3.0  &  0.32  &  1.05  &  1.37  \\
		 & 10.0  &  0.05  &  0.25  &  0.30   \\
 200000  &  1.0  &  0.23  &  0.21  &  0.44   \\
         &  3.0  &  0.04  &  0.21  &  0.25   \\
		 & 10.0  &  0.01  &  0.04  &  0.05  \\
\hline

\end{tabular}
\end{table}

\section[]{The Landau-Pomeranchuk-Migdal effect}

In both this and previous work, we have adopted  expressions
for the zero-pressure ion number density,
\begin{eqnarray}
N = 2.5\times 10^{28}Z^{-0.7}B^{1.2} \hspace{3mm}
{\rm cm}^{-3},
\end{eqnarray}
and for the interatomic separation in a linear chain,
parallel with ${\bf B}$,
\begin{eqnarray}
a_{s} \approx 1.5\times 10^{-10}Z^{1/2}B^{-1/2}\hspace{3mm}
{\rm cm},
\end{eqnarray}
conveniently summarizing the calculated values of
Medin \& Lai (2006).
We anticipate that the bremsstrahlung and pair creation cross-sections
obtained in previous Sections for isolated atoms 
will be considerably modified in the
condensed matter of the neutron-star surface. In particular,
it is essential to consider the Landau-Pomeranchuk-Migdal (LPM)
effect, whose existence at zero field has been established
without doubt by experiments with laboratory electron beams of
several hundred GeV (Hansen et al 2004). The essence of the
effect is that, for example in bremsstrahlung and at high
energies, the isolated-atom longitudinal coherence length (of the 
order of $q^{-1}$)
can be so long that the coherence that would be present in
the case of an isolated atom is partially removed by
multiple scattering of the initial or final-state electron
within the medium.  We refer to
the review by Klein (1999) for further details, and for a
discussion of other processes in condensed matter, such as the
dielectric effect, that are not relevant here.  In the present
Section, we examine the LPM effect at high magnetic fields and find
that the one-dimensional nature of kinematics at $B \sim B_{c}$
allows a particularly simple description.

We begin by considering the effect of scattering of an electron
with initial longitudinal momentum $p \gg mc$ by a (dimensionless)
potential $U^{i} = -Ze^{2}\tilde{V}/\hbar c$
whose origin is at $z = z_{i}$.  We consider transitions with
no change of Landau quantum number so that the final state
is given by the Green function,
\begin{eqnarray}
G = -i\theta(t - t^{\prime})\sum _{p^{\prime}>0}\psi _{n}
({\bf r}_{\perp})\psi _{n}^{*}
({\bf r}^{\prime}_{\perp}){\rm e}^
{ip^{\prime}(z-z^{\prime})-iE^{\prime}(t-t^{\prime})},
\end{eqnarray}
constructed from the Landau functions given by equation (A3).
The truncation to $\delta n = 0$ is possible because, as we found
in Section 2, the $\delta n \neq 0$ transition rates are at
least an order of magnitude smaller.  It also excludes
spin-flip and back-scattering $(p^{\prime} < 0)$ which are both
negligible in the present context.  The scattered state is
\begin{eqnarray}
\int d^{3}r^{\prime}dt^{\prime} G({\bf r},t;
{\bf r}^{\prime},t^{\prime})U^{i}\psi _{n}({\bf r}^{\prime}_{\perp})
{\rm e}^
{ipz^{\prime}-iEt^{\prime}} =   \nonumber    \\
\int^{\infty}_{-\infty} \frac{dq}{2\pi}\frac{U^{i}_{q}}
{-q + i\eta}{\rm e}^{iq(z - z_{i})}
\left(\psi _{n}({\bf r}_{\perp}){\rm e}^{ipz
-iEt}\right),
\end{eqnarray}
in which $U^{i}_{q}$ is the matrix element of $U^{i}$
between states $p$ and $p^{\prime} = p + q$,
and $\eta > 0$ is infinitesimal. It is
reached by introducing the integral representation of the Heaviside
function $\theta$, followed by integration over ${\bf r^{\prime}}$ and 
$t^{\prime}$.  Equation (17)
shows that the electron has components of momentum
$p + q$ and that
these are significant at distances $z- z_{i} < q^{-1}$.  Either by
performing the integral, or taking the limit $z\rightarrow\infty$,
we can see that the prefactor outside the brackets evolves to an
amplitude $U^{i}_{0}$, as expected.  Apart from the
forward-scattered component of the wave-function given by
equation (17), flux conservation shows that there is also an
undisturbed component of amplitude $(1 - |U^{i}_{0}|^{2})^{1/2}$.
Equation (17), giving the forward-scattered state, is no more than
a concrete expression of the uncertainty
principle, but it shows that forward scattering by a sequence
of atoms, though unobservable in the motion of the electron
itself, has the effect of broadening
the distribution of $q$ and so reducing the small-$q$ amplitudes
that mediate most of the bremsstrahlung cross-section at
high energies. 

The simplicity of
equation (17) enables us to find the Green function for an electron
undergoing a sequence of forward scatters and so provides an easy
way to calculate bremsstrahlung cross sections including the
LPM effect.  We begin by noting that the distribution in the
variable $q$ in equation (17) is unchanged in form
by multiple forward scattering as is shown by the amplitude for
a second and successive scattering at
$z_{j}$,
\begin{eqnarray}
\int^{\infty}_{-\infty}
\frac{du}{2\pi}\int^{\infty}_{-\infty}
\frac{dq^{\prime}}{2\pi} 
 \frac{{\rm e}^{iu
(z-z_{j})}}{-u+i\eta}U^{j}_{u}
\frac{{\rm e}^{iq^{\prime}(z-z_{i})}}{-q^{\prime}+i\eta}
U^{i}_{q^{\prime}} =     \nonumber   \\
i\int^{\infty}_{-\infty}\frac{dq}{2\pi}U^{j}_{q}U^{i}_{0}
\frac{{\rm e}^{iq(z-z_{j})}}{-q+i\eta},
\end{eqnarray}
in which $u = q - q^{\prime}$.  We assume that the properties of
$U$ allow the limits
of integration to be extended to $\pm\infty$ with completion of
the contour in the upper half of the complex-$q^{\prime}$ plane.
We see that the amplitude has the spatial dependence $z-z_{j}$ of
the last scattering.

Consequently, the amplitude for an electron propagating
without back-scattering or $\delta n \neq 0$ transitions
in a long interval of $z$ must be a superposition
of $q$-distributions, each of the form of equation (17), with
components which are necessarily identical except that their
individual amplitudes decay exponentially as functions of
$z-z_{j}$, reflecting a reduction factor 
$(1 - \langle|U_{0}|^{2}\rangle)^{1/2}$ for each atom traversed.
This exponential representation is satisfactory
provided the mean square value of $|U^{j}_{0}|^{2}$, satisfies
the condition $\langle|U_{0}|^{2}\rangle \ll 1$, as it does here.
Therefore, we can find the Green function that couples the
electron to a final state of total
longitudinal momentum $K_{\parallel}$.  It is given by, 
\begin{eqnarray}
\sum _{j}\int^{\infty}_{z_{j}}dz{\rm e}^{-iK_{\parallel}z}
\int^{\infty}_{-\infty}\frac{dq}{2\pi}\frac{U^{j}_{q}}{-q+i\eta}
{\rm e}^{ipz + iq(z-z_{j}) - (z-z_{j})/2\lambda} \nonumber \\
 = \sum _{j}\left(\frac{-1}
 {p - K_{\parallel} + i/2\lambda}\right)
 U^{j}_{0}{\rm e}^{i(p - K_{\parallel})z_{j}},
\end{eqnarray}
in which $\lambda = \langle a\rangle/\langle|U_{0}|^{2}\rangle$,
where $\langle a\rangle$ is the mean interatomic separation
$z_{j} - z_{j-1}$ and we have
integrated first over $z$ and then over $q$ by completion of
the contour in the semi-infinite upper half of the complex
$q$-plane.  The right-hand side of equation (19) contains,
within brackets, the
momentum-space Green function for an electron propagating
without back-scattering, spin-flip or $\delta n \neq 0$ transitions.
Forming the square modulus of the remaining term (and
neglecting any inhomogeneity in the $U^{j}_{0}$) yields
the (one-dimensional) static structure function
for a linear chain of $N_{a}$ atoms,
\begin{eqnarray}
S(q) = \frac{1}{N_{a}}\left|\sum _{j=1}^{N_{a}}
{\rm e}^{iqz_{j}}\right|,
\end{eqnarray}
which contains, as a function of $q=K_{\parallel}-p$,
the dependence of the transition rate on the
condensed-matter structure.
Allowance for transitions with change of Landau quantum number
should not be difficult but is not essential because, as can
be seen from the Table 1 cross-sections, their mean free path
is considerably longer than $\lambda$.

Pair production and bremsstrahlung cross-sections including the
LPM effect are obtained
immediately, assuming a random distribution of atoms defined by
a structure function $S(q) = 1$,
simply by substituting this Green function for the
energy denominators contained in equations (8) or (11).  We
have done so and have repeated
the calculations by which the Table 3 and 4 cross-sections were
obtained.
The revised cross-sections given in the lower sectors of these
Tables are for $Z = 26$ and require some comment. 

In the case of pair production, the re-calculated cross-sections
are given in Table 3 just for $B = B_{c}$ and show that, owing
to the very
high matter density ($\sim 2\times 10^{5}$ g cm$^{-3}$) the LPM
effect becomes noticeable at $k = 4000$ and reduces
cross-sections by more than an order of magnitude at $k = 40000$.
However, its effect on bremsstrahlung is less simple.  The
re-calculated cross-sections in Table 4 show that there is
an initial increase at low values of $p$.  The explanation for
this is as follows.  For fixed values of the kinematic variables
and quantum numbers, there is a partial cancellation of the two
terms in equation (11) which arises because the energy
denominators are almost exactly equal in magnitude
($\approx \pm q $) but have opposite signs.
The presence of $\lambda^{-1}$ in the modified Green function
creates a further amplitude term in which this cancellation is
not present, so producing the cross-section increase.  But with
increasing $p$, values of $q$ decline and eventually the
$\lambda^{-1}$ component of the denominator becomes dominant.
Its effect becomes marked for $p > 20000$ mc.

The presence of the
LPM effect removes the simple $Z^{2}$ scaling of cross-sections,
but equations (14) and (15) allow us to see that
$\lambda \propto Z^{\beta}$, with $-1.77 < \beta < -1.5$.  Thus for
a fixed magnetic field,
cross-sections become asymptotically dependent on $Z^{2+2\beta}$
in the high-energy limit.

But the possibility that zero-pressure neutron-star matter has some
degree of one-dimensional order must also be considered.  Formation
of one-dimensional chains of atoms of homogeneous $Z$, oriented
parallel with the magnetic flux, is energetically favoured 
at $B\sim B_{c}$ (see Medin \& Lai 2006) and
such structure may be present at the surface to the extent that it
survives the formation of the approximately Poisson distribution
of nuclear charge produced  by
photodisintegration reactions in
the reverse-electron electromagnetic showers.
Its effect can be studied simply by reference to the
structure factor given by equation (21) neglecting, for the moment,
any inhomogeneity in atomic number.  Consider, for example, the
structure function for a finite linear chain of $N_{a}$
homogeneous atoms with spacing $z_{j}-z_{j-1} = a_{s}$,
\begin{eqnarray}
S(q) = \frac{1}{N_{a}}\left(\frac{\sin(N_{a}qa_{s}/2)}
{\sin(qa_{s}/2)}\right)^{2}.
\end{eqnarray}
This has maxima, $S(q) = N_{a}$, when $q$ coincides with an integral
multiple of the basic reciprocal-lattice wavenumber $g = 2\pi/a_{s}$,
but has zeros at $|q -g| = g/N_{a}$.  Its effect is primarily to
significantly distort spectral shapes given by
isolated-atom differential
cross-sections.  Unfortunately, it is not easy to be specific about
changes in total cross-sections  because they are dependent on the
(unknown) structure function and on all other parameters,
but with a change of variable in a particular differential
cross-section to $q$ we can note that the modified total
cross-section is,
\begin{eqnarray}
\sigma^{S} = \int S(q)\frac{d\sigma}{dq}dq,
\end{eqnarray}
taken over the complete interval of $q$ for the process, and that
this differs from the $S = 1$ cross-section by a factor of order
unity, not $N_{a}$.

\section[]{Conclusions}

The cross-sections obtained here are those that, with Compton
scattering, determine electromagnetic shower development.  For
magnetic flux densities of the order of $10^{12}$ G, Landau
quantum numbers $n \sim 10^{2}$ are associated with transverse
momenta $p_{\perp} \sim mc$.  Thus, on the basis of the
correspondence principle, zero-field expressions are used with
some degree of confidence.  But in many neutron stars, fields more
than two orders of magnitude larger have been inferred from the
observed spin-down rate.  For this reason, it is necessary to be
aware of any field-dependence of these cross-sections that would
qualitatively change the nature of shower development.  In this
respect, the processes that appear not to have been considered
previously are those of Sections 2, 4 and 5.  Of these, 
Rutherford scattering merits little comment except
to note that its cross-section decreases with increasing magnetic
field in the region $B \sim B_{c}$.  Pair creation is a little
more complex because the amplitude (equation 8) becomes singular
at the threshold values of $k_{\perp}$ for magnetic pair creation,
beyond which it becomes no more than a Coulomb field-dependent
correction to that process.  Thus we are concerned with the
cross-section for $k_{\perp}$ below the first two thresholds
discussed following equation (6) in Section 3. This is the
explanation for the intervals of $k_{\perp}$
adopted for columns 3-5 of Table 3.

Bremsstrahlung at $B \sim B_{c}$ is an even more complex process.
It is of second order in the coupling of the electron with the
electromagnetic field, as at zero field. But there are two significant
differences.  The Table 4 cross-sections are for an electron
initially in the $n = 0$ Landau ground state, but in the
development of a shower in condensed matter, higher-$n$ states
are certainly populated owing to the very large Rutherford
cross-sections listed in Table 1.
Secondly, a pole in the amplitude (equation 11) for such
initial states represents
the possibility of real intermediate states of $n^{\prime} \geq 1$.
The process we refer to as bremsstrahlung then consists of two
separate first-order processes; Rutherford scattering followed by
cyclotron
emission with the natural widths given in Table 2.  The initial
electron momentum determines whether or not this is more important
than the second-order process. An estimate of the critical value
is easily made from the cyclotron emission rates in
Table 2 which are all of the order of $10^{-3}\omega _{B}$. Although
the mean free path $l_{Ru}$ for Rutherford scattering is always
small compared with the radiation length $l_{Rad}$ which
is characteristic of the second-order bremsstrahlung process,
time dilation can extend the
mean flight path $l_{ce}$ for natural decay by cyclotron emission
to lengths greater than $l_{Rad}$.  
Then the first-order processes are the dominant source of photon
production provided the initial-state momentum is,
\begin{eqnarray}
\frac{p}{mc} < \frac{l_{Rad}}{c}\sum _{n^\prime=0}^{n-1}
\Gamma _{nn^{\prime}}.
\end{eqnarray}
From the transition rates given in Table 2 and the values of
$\sigma^{Rad}$ given in Table 4, we can see
that its value is not strongly $B$-dependent.  Specifically,
for $B = B_{c}$, $Z = 26$, $\sigma^{Rad} = 22$ bn and
$l_{Rad} = 1.8\times 10^{-5}$ cm with inclusion of the LPM
effect, so that for
$n = 4$, the inequality (23) becomes $p < 1100$ mc. 
At lower momenta, the process of bremsstrahlung becomes simply
a sequence of excitation (or de-excitation) of higher-$n$
Landau states by Rutherford scattering and de-excitation by
cyclotron
emission.  At even lower momenta, the balance between excitation
and de-excitation
changes and the typical values of the Landau quantum
number decrease until for $l_{Ru} \approx l_{ce}$, that is, for
\begin{eqnarray}
\frac{p}{mc} \approx \frac{l_{Ru}}{c}\Gamma _{10}
\end{eqnarray}
they are $n = 1$.  With the cross-section $\sigma _{01}$ from
Table 1, we find $l_{Ru} =4.0\times 10^{-7}$ cm.  The transition
rate $\Gamma _{10} = 1.06 \times 10^{-3}\omega _{B}$ gives
$p \approx 11$ mc.

The above summary of the properties of the bremsstrahlung and
pair creation processes at high magnetic fields serves to show
qualitatively how the characteristics of electromagnetic showers
differ from the zero-field case.  In the initial stages of an
electron-initiated shower of about $10^{3}$ GeV, the
Landau-Pomeranchuk-Migdal (LPM) effect is far more effective
in reducing cross-sections owing to the very high density of
the condensed matter at the neutron star surface.  Therefore,
shower development is displaced inward until typical electron
or photon energies are reduced to the extent that the effect
becomes unimportant.  At this stage, electron energy loss becomes
rapid, occurring within lengths much shorter than $l_{Rad}$.
Photon mean free paths are dependent on $k_{\perp}$, but
below the lowest threshold for magnetic pair creation at
$k_{\perp} < 2$ they are a little larger than the value of
given by the zero-field Bethe-Heitler cross-section.

The principal interest in shower development here is in proton
production by decay of the nuclear giant dipole resonance (GDR).
Cross-sections for its electro-production are small so that the
important shower property is the distribution of total photon
track length in the neighbourhood of the resonant momentum,
$k \approx 41$ mc.  An estimate of this, made under the
assumptions that  bremsstrahlung occurs through the sequence of
first-order processes decribed above and that the Bethe-Heitler
pair creation cross-section is valid, was given by Jones (2010).
We have found that shower development at $B \sim B_{c}$ is
more complex than was assumed in that work, the important
question being the fraction of photon production in the GDR
region that has transverse momentum below the
magnetic pair creation thresholds at $k_{\perp} = 2$ for Landau
quantum numbers $n_{+} = n_{-} = 0$ and $k_{\perp} = 1 + \sqrt{1+2B}$
for $n_{\pm} = 0$ with $n_{\mp} = 1$.

It is significant that cyclotron decay, which is dominant in the
later stages of shower development, always leads to a decrease in
Landau quantum number, so bringing about a convergence of the shower
toward small $k_{\perp}$.  The only qualification here is the
tendency, noted by Harding \& Lai (2006) and confirmed in Table 2,
for partial transition rates direct to the ground state
$n = 0$ to be large at $B \gg B_{c}$.  Complete and quantitative
calculations
of shower development would require an expansion of Tables 1, 2 and
 4 to much larger initial-electron values of the Landau quantum number,
and a Monte Carlo calculation using them would be a substantial
undertaking which has not been attempted here.  Qualitatively, we
can conclude that the photon track length distribution assumed in
previous work is valid at $B \approx B_{c}$.
For very high fields at $B \gg B_{c}$, the pair creation
cross-sections become small and the photon mean free path at
transverse momenta below magnetic pair creation thresholds are
then determined by Compton scattering. Therefore,
we expect that the effect of  the large cyclotron partial transition
rates to $n = 0$ will begin to reduce total photon track length 
in the GDR region of momentum at $B \gg B_{c}$.

\appendix
\section[]{Solutions of the Dirac equation}

The solutions used here are the spinors obtained
by Johnson \& Lippmann (1949)
for a free electron in a uniform magnetic field with the gauge
${\bf A} = ({\bf B}\times{\bf r})/2$.  As these authors noted,
they can be easily expressed
in terms of any complete set of eigenfunctions of the
non-relativistic Schrodinger Hamiltonian, whether in cartesian or
polar coordinates.  The Dirac equation
is,
\begin{eqnarray}
\left(\mbox{\boldmath$\alpha$}\cdot\mbox{\boldmath$\pi$} +
 \beta m\right)\Psi =
 E\Psi,
\end{eqnarray}
in which $\mbox{\boldmath$\pi$} = {\bf p}+e{\bf A}$
with $e > 0$ and $c = 1$.
The solutions for $E > 0$ and for Landau quantum number $n \geq 0$ are,
\begin{eqnarray*}
\Psi_{-1,n} = \left[\begin{array}{c}
0  \\
(m+E)\psi _{n} \\
p_{\perp}\psi _{n-1}  \\
-p\psi _{n}  
\end{array} \right] \hspace{5mm}
\Psi_{1,n-1} = \left[\begin{array}{c}
(m+E)\psi _{n-1} \\
0  \\
p\psi _{n-1}  \\
p_{\perp}\psi _{n}
\end{array}  \right]
\end{eqnarray*}
each multiplied by the common normalization factor
$G = 1/\sqrt{2E(E+m)}$. 
The longitudinal momentum component is $p$.  These solutions have
eigenvalues given by equation (1) and are
two-fold degenerate for $n > 0$, but they are not identical with the
true physical states owing to radiative corrections that have been
neglected. 
Natural line-widths, and level shifts arising from the electron
anomalous magnetic moment and from radiative corrections, are all
of similar magnitude and give physical states that are quite
different from the above solutions (see Herold, Ruder \&
Wunner 1982; also Baring, Gonthier \& Harding 2005).  The
distinction would be important if spin-polarized electron or
positron states were being considered.  However, for the
cross-sections obtained here, which are for relativistic
energies, lepton spin-polarization is not of interest and the 
fine-structure in energy levels can be safely neglected. In 
this paper, we always average or sum over lepton spin states
so that we require only the completeness of the Johnson-Lippmann
states.  These, with the negative energy solutions given by the
transformation $E \rightarrow -E$, have been used explicitly in
Sections 2 - 5. The corresponding
representation of the Dirac matrices is easily found from equations
(47)-(54) of Johnson \& Lippmann and is
\begin{eqnarray*}
\mbox{\boldmath$\alpha$} = \left(\begin{array}{cc}
0 & \mbox{\boldmath$\sigma$}\\
\mbox{\boldmath$\sigma$} & 0
\end{array} \right) \hspace{1cm}
\beta = \left(\begin{array}{cc}
I & 0 \\
0 & -I
\end{array} \right).
\end{eqnarray*}
The subscript $s=\pm 1$ labelling the spinors indicates that for the case
$n = 0$, or in the limit $B \rightarrow 0$, they are also eigenstates of
$\sigma _{z}$ with eigenvalues as indicated.
The functions $\psi _{n}$ are a complete orthonormal set of eigenfunctions of 
$\mbox{\boldmath$\pi$}^{2}$ labelled by the Landau quantum number $n$.  Their 
relative phases are not arbitrary because they are required to satisfy the 
conditions,
\begin{eqnarray}
(\pi _{x} + i\pi _{y})\psi _{n-1}  =  p_{\perp}(n)\psi _{n}, \nonumber \\
(\pi _{x} - i\pi _{y})\psi _{n} =  p_{\perp}(n)\psi _{n-1},
\end{eqnarray}
where $p_{\perp}(n) = \sqrt{2nB}$ and B is the magnetic flux density in units of 
$B_{c}$.  The atomic Coulomb potential has axial symmetry, reflecting the shape 
of the electron orbitals in the magnetic field. (We
neglect any small component with a lattice symmetry.)
Thus cylindrical polar coordinates $r_{\perp}$, $\phi$, $z$, with ${\bf B}$ 
parallel with the $z-axis$ appear to be the optimum.  The functions are then,
\begin{eqnarray}
\psi _{nl} = \frac{(-i)^{n}}{r_{B}\sqrt{n!l!}}\xi^{(l-n)/2}{\rm e}^{-\xi/2}
f_{n}(l,\xi)
\frac{{\rm e}^{-i(l-n)\phi}}{\sqrt{2\pi}}{\rm e}^{ipz},
\end{eqnarray}
conveniently expressed in terms of the variable
 $\xi = r_{\perp}^{2}/2r_{B}^{2}$, where $r_{B} = mc^{2}\sqrt{B}/e$
is the cyclotron radius.
The Landau quantum number can be any positive integer $n = 0, 1, 2, ..$
and all except the $n = 0$ state are two-fold degenerate.
In addition, the quantum number $l$ represents
the infinite spatial degeneracy of all Landau states whose
guiding centres can be at radii $\sqrt{2l + 1}r_{B}$ from the z-axis.
The polynomials
$f_{n}$ are,
\begin{eqnarray}
f_{0} & = & 1    \nonumber  \\
f_{1} & = & l - \xi  \nonumber   \\
f_{2} & = & l(l-1) - 2l\xi + \xi^{2}  \nonumber   \\
f_{3} & = & l(l-1)(l-2) - 3l(l-1)\xi + 3l\xi^{2} - \xi^{3} \nonumber  \\
f_{4} & = & l(l-1)(l-2)(l-3) -4l(l-1)(l-2)\xi  \nonumber \\ 
   &  & \mbox{}  + 6l(l-1)\xi^{2} - 4l\xi^{3} + \xi^{4}
\end{eqnarray}
in obvious sequence and they satisfy,
\begin{eqnarray}
f_{n}(l,\xi) = (-\xi)^{n-l}f_{l}(n,\xi),
\end{eqnarray}
which is worth noting in relation to states with $l < n$.

We adopt gaussian cgs units so that the fine structure constant is
$e^{2}/\hbar c$.  Perturbations of the correct sign are introduced into
equation (A1) by modifying the energy-momentum four-vector,
${\bf A}\rightarrow {\bf A} + \tilde{\bf A}$ and
$E\rightarrow E - Ze^{2}\tilde{V}$, to include the radiation field
$\tilde{\bf A}$ and the atomic Coulomb potential for nuclear charge $Z$.
Thus the absorption of a photon with wave-vector ${\bf k}$
and angular frequency $\omega$ is represented by the matrix element of
\begin{eqnarray}
e\mbox{\boldmath$\alpha$}\cdot\tilde{\bf A} = \left(\frac{2\pi\hbar c^{2}}
{\omega}\right)^{1/2}e\mbox{\boldmath$\alpha$}\cdot
\mbox{\boldmath$\epsilon$}{\rm e}^{i{\bf k}\cdot{\bf r}},
\end{eqnarray}
in which $\mbox{\boldmath$\epsilon$}$ is the polarization vector
either
parallel with, or perpendicular to, the vector
${\bf k}\times{\bf B}$.  

The creation of a pair in states specified by
$|p_{\pm},n_{\pm},l_{\pm},s_{\pm}\rangle$, as in lines three and six
of equation (8), 
is given by the matrix element
\begin{eqnarray}
    &  &  \int r_{\perp}dr_{\perp}d\phi dz 
	\Psi _{s_{-},n_{-}}^{\dagger}(p_{-}, E_{-}, l_{-})
e\mbox{\boldmath$\alpha$}\cdot\tilde{\bf A}  \nonumber    \\
    &  &  \hspace{2cm} \Psi _{-s_{+},n_{+}}(-p_{+}, -E_{+}, l_{+}),
\end{eqnarray}
in which the positron energy is $E_{+} > 0$.
Evaluation of this proceeds by matrix
multiplication and immediately yields, for each set of values of
$s_{\pm}$ and of photon polarization, a linear combination of terms
each of which,
apart from kinematic factors, contains an integral of the form,
\begin{eqnarray*}
\lefteqn{I(k_{\perp},n_{-},l_{-},n_{+},l_{+}) =} \hspace{5cm} \\
 \int r_{\perp}dr_{\perp}d\phi
{\rm e}^{ik_{\perp}r_{\perp}\cos\phi}\psi^{*}_{n_{-},l_{-}}({\bf r}_{\perp})
\psi _{n_{+},l_{+}}({\bf r}_{\perp}).
\end{eqnarray*}
These expressions, and the integrals contained in them, have all been
evaluated numerically.  Therefore, we are unable to give useful
analytical expressions for any of the cross-sections calculated in this
work. Evaluation is the more tedious because the transverse dipole
approximation is not adequate so that there is no radiative transition
selection rule that is effective for physically significant values of 
$k_{\perp}$. But matrix elements of the Coulomb potential satisfy
$\delta(n - l) =0$ strictly for isolated atoms; also in condensed
matter to the extent that components with the symmetry of the
crystalline field are small.

\bsp

\label{lastpage}

\end{document}